%Paper: hep-ph/9307296
%From: "Judy Mack, University of Rochester, 716-275-4840"
%Date: Mon, 19 Jul 1993 11:28 EST

\magnification=1200
\baselineskip=13pt
\input ABCmacro
\input tables
\rightline{UR-1314\ \ \ \ \ \ \ }
\rightline{ER-40685-764}
\ \ \ \ \
\vskip 1.5cm
\baselineskip=20pt
\centerline{\bf TWO-BODY NONLEPTONIC DECAYS OF CHARMED MESONS}
\bigskip

\bigskip
\bigskip

\centerline{P. Bedaque, A. Das and V. S. Mathur}
\centerline{Department of Physics and Astronomy}
\centerline{University of Rochester}
\centerline{Rochester, NY 14627}

\vskip 2cm

\noindent {\bf \underbar{Abstract}}

Two-body nonleptonic decays of charmed mesons are studied on the basis of
 a simple pole-dominance model involving the vector, pseudoscalar and
axial-vector meson poles.

\vfill\eject
\noindent {\bf 1. \underbar{Introduction}}
\smallskip

In this paper, we calculate the two-body non-leptonic decays of charmed
mesons using the pole-dominance model.  A preliminary analysis of this type
was first attempted$^1$ several years ago, but we believe that a systematic
investigation of the type presented here has not been carried out.  The
 non-leptonic weak decay of a pseudoscalar meson into two pseudoscalar mesons
is parity violating, so in such decays the pole-dominace model would
involve only vector-meson poles.  It is well-known$^2$ that this idea of vector
dominance provides a successful description of the $K \rightarrow \pi \pi$
decays.  Recently we have also discussed$^3$ the decay of $D$ mesons into two
pseudoscalars ($PP$) through vector dominance and there, too, we find
reasonable agreement with experiments.  In the present work we extend
the analysis to the general two-body non-leptonic decays of $D$ mesons to
include the decay to a pseudoscalar and a vector ($PV$) meson and to two
vector ($VV$) mesons.  Since we now deal with parity conserving decays as well,
we have to go beyond vector dominance.  In this work, we extend our
consideration to include also the pseudoscalar and the axial vector-meson
poles.

Our work has no direct relationship to the extensively studied
factorization model$^4$ which has generally been quite successful in describing
the decays except for those that proceed only through what are known as the
annihilation diagrams.  The pole-dominance model, on the other hand, has
enjoyed many successes in low-energy hadron phenomenology, and it would be
interesting to see how it describes charm decay.$^5$

\medskip

\noindent {\bf 2. \underbar{Preliminaries}}

\smallskip

For non-leptonic decay of charm, the effective weak Hamiltonian may be
written in the current-current form as$^6$
$$H_W = {G_F \over \sqrt{2}} \ \left[ a_1 (\overline u d^\prime )_\mu
(\overline s^\prime c)_\mu + a_2 (\overline s^\prime d^\prime )_\mu
(\overline uc)_\mu \right] \eqno(1)$$
where $\left( \overline q^\alpha q_\beta \right)_\mu$ are color-singlet V-A
currents
$$\left( \overline q^\alpha q_\beta \right)_\mu = i\  \overline q^\alpha\
 \gamma_\mu \left( 1 + \gamma_5 \right) q_\beta = \left( V_\mu
\right)^\alpha_\beta + \left( A_\mu \right)^\alpha_\beta \eqno(2)$$
\noindent
The indices $\alpha,\beta=1,2,3,4$ represent the flavour $SU(4)$ indices
and $a_1,\ a_2$ are real coefficients which we treat as phenomenological
parameters.  The primed quark fields are related to the unprimed ones by
the usual Cabibbo-Kobayashi-Maskawa (CKM) mixing matrix.  For the
non-leptonic decays of $D$ mesons into two mesons, the Hamiltonian (1)
leads to two main classes of quark-model diagrams, the spectator and the
annihilation diagrams shown in Figs. 1(a) and 1(b), respectively.  We
ignore the Penguin-type contributions.  It is well-known that the
annihilation-diagram contribution in the quark model is helicity
suppressed.

In the pole-dominance model, we take the currents in $H_W$ to be the
hadronic currents given by the field current identities $(\alpha ,
 \beta = 1,2 \dots 4)$
\smallskip
$$\eqalign{\big( V_\mu \big)^\alpha_\beta &= \sqrt{2}\ g_V \big( \phi_\mu
\big)^\alpha_\beta \cr
\noalign{\vskip 4pt}%
\big( A_\mu \big)^\alpha_\beta &= \sqrt{2}\ f_P \partial_\mu
P^\alpha_\beta + \sqrt{2}\ g_A \big( a_\mu \big)^\alpha_\beta \cr}
\eqno(3)$$
\smallskip
\noindent where
 $\left( \phi_\mu \right)^\alpha_\beta$, $P^\alpha_\beta$ and $\left(
a_\mu \right)^\alpha_\beta$ are the field operators of the vector, the
pseudoscalar and the axial-vector mesons of $SU(4)$, respectively, and
 $g_V,\ f_P\
{\rm and}\ g_A$ are the corresponding decay constants.  The non-leptonic weak
interaction can then be represented by a two-meson vertex which can be read
off from (1) upon substituting (3).

\medskip

\noindent {\bf 3. \underbar{Calculation of the Decay Amplitude}}

\smallskip

The Feynman diagrams for the two-body decays $D \rightarrow PP,\  PV$ and
$VV$ are displayed in Figs. 2-4.  In these figures, the dark dot represents a
weak interaction vertex and the open circle a strong vertex.  Also the dotted
lines, solid lines and wavy lines represent pseudoscalar, vector and
axial-vector mesons respectively.  Note that the figures (2b), (3c), (3e),
(4b) and (4c) are the analogues of the annihilation diagrams in the quark
model, the remaining figures corresponding to the spectator diagrams.
Only the vector-meson pole contributes to the parity violating decay $D
 \rightarrow PP$ as shown in Fig. 2.  Also, since $D \rightarrow PV$
 is a parity conserving decay, a diagram like Fig. (3b) where the
pseudoscalar pole is replaced by a vector pole cannot contribute.

We have several types of strong vertices appearing in the diagrams.  These
include the $VPP$, $VVP,$ $VVV$ and the $VPA$ vertices.  Unfortunately,
many of these couplings are not known.  For numerical work, we choose to
relate these couplings by a suitable flavor symmetry.  In fact the
$VVP$ and $VVV$ couplings can be related to the $VPP$ couplings through an
extended spin-SU(4) symmetry.  Accordingly, we take the strong
Hamiltonian as
$$\eqalign{H_{\rm str} = i\ g\ {\rm Tr}\ \big[ &\phi_\mu P
{\buildrel \longleftrightarrow \over{\partial_\mu}}
 P - {2 \over M}\ \varepsilon_{\mu \nu
\lambda \rho} P \partial_\mu \phi_\nu \partial_\lambda
\phi_\rho\cr
\noalign{\vskip 4pt}%
&+ {2 \over 3}\ F_{\mu \nu} \phi_\mu \phi_\nu - {2 \over
9M^2}\ F_{\mu \nu} F_{\nu \lambda} F_{\lambda \mu}
\big] \cr}\eqno(4)$$
This is an obvious generalization of the Sakita-Wali interaction
Hamiltonian$^7$ which is relevant to flavor SU(3).  In (4), the trace is over
the SU(4) multiplets, $g$ is the coupling constant and $M$ represents a
mass scale.  We shall identify $M$ with the mass of the decaying particle,
and take $g$ to be the $\rho \pi \pi$ coupling determined from the
$\rho$-width.  The $VPA$ interaction is not contained in (4) and will be
taken to be$^8$
$$H_{\rm str} (VPA) = i\ g_s \ M\ {\rm Tr}\
\left\{ \phi_\mu [P , a_\mu] \right\} +
{ig_d \over M}\ {\rm Tr}\ \left\{ \partial_\mu \phi_\nu
[P, \partial_\nu a_\mu ] \right\} \eqno(5)$$
Despite some attempts in the past,$^{8,9}$
 the coupling constants $g_s$ and
 $g_d$ have not been determined successfully.  In our present analysis, we
will treat these as unknown parameters.

It should be mentioned that experience based on flavor SU(3) symmetry
indicates that coupling constants are better described by
group symmetry than the masses.
Furthermore, the vector-meson couplings may be expected to
have a somewhat special status embodied through universality.  At the
present time, however, we can only hope that couplings related by the
larger SU(4) symmetry are not unreasonable.  Future experiments involving
the strong decays of charmed vector or axial vector mesons would shed light
on this issue.

The weak vertices in Figs. 2-4 involve the decay constants $g_V,$ $f_P$ and
$g_A$ for various particles.
  Once again only a few of these are known from experiments.  Now, spectral
function sum-rules$^{10,11}$
 based on asymptotic flavor symmetries have been quite
successful in the past, and we may use them here to obtain suitable
relations.  From asymptotic SU(4) symmetry, it is easy to derive$^{12}$
 the result
that $g_V / m_V$ would be the same for all vector mesons.  Since we can
extract $g_\rho$ from the data on the decay $\rho \rightarrow \ell
\ \overline \ell$, this result leads to a determination of all $g_V$.
Also from asymptotic chiral SU(4) symmetry, we can derive the spectral function
sum-rule$^{13}$
$$g^2_A / m^2_A + f^2_P = g^2_V / m^2_V \eqno(6)$$
where $A, \ P\ {\rm and}\ V$ are particles with the same internal quantum
numbrs.  Now experimentally, only $f_\pi$ and $f_K$ are known.  However,
$f_D$ has
  been determined$^{14}$ from QCD sum-rules.$^{15}$
  With the $g_V$'s and
$f_P$'s known, the sum-rule (6) then serves to determine the
$g_A$'s.

The numerical values of the various couplings used in our analysis are
displayed in Table 1.  For the axial vector meson multiplet, we have taken
the particles to be$^{16}$
 $A_1(1260),$ $K_1 (1400),$ $f_1 (1285)$, $D_1 (2420)$
and $D_{S_1} (2536)$.

The decay amplitudes for the $PP$, $VP$ and $VV$ modes are defined as
follows
$$\eqalignno{&M (D (q) \rightarrow P_1 (q_1) P_2 (q_2)) = {-i (2\pi)^4
\delta^{(4)} (q-q_1 - q_2) \over
\sqrt{2 q_0 V 2q_{10} V 2 q_{20} V}} \ iA &(7)\cr
\noalign{\vskip 4pt}%
&M(D(q) \rightarrow V_1 (q_1, \lambda_1 ) P_2 (q_2)) =
{-i (2 \pi)^4 \delta^{(4)} (q - q_1 - q_2 )   \over
\sqrt{2 q_0 V 2q_{10} V 2 q_{20} V}} \ q_2 \cdot
\varepsilon^{(\lambda_1)} (q_1) \ B  &(8)\cr
\noalign{\vskip 4pt}%
&M(D(q) \rightarrow V_1 (q_1, \lambda_1 ) V_2 (q_2,
 \lambda_2)) =
{-i (2 \pi)^4 \delta^{(4)} (q - q_1 - q_2 )   \over
\sqrt{2 q_0 V 2q_{10} V 2 q_{20} V}}\cr
\noalign{\vskip 4pt}%
&\qquad \big[ i \ C\ \delta_{\alpha \beta} + i\ D \
\varepsilon_{\mu \alpha \nu \beta}
q_{1 \mu} q_{2 \nu} + i\ E\ q_{1 \beta} q_{2 \alpha}\big]
\varepsilon_\alpha^{(\lambda_1)} (q_1)
\varepsilon^{(\lambda_2)}_\beta (q_2) &(9)\cr}$$
where $A,\ B \dots E$ are constants representing the invariant amplitudes.
The decay widths are easily calculated in terms of these amplitudes to be
$$\eqalignno{\Gamma (D \rightarrow P_1 P_2) &= {1 \over 8 \pi} \
{k \over M^2}\ |A|^2 &(10)\cr
\noalign{\vskip 4pt}%
\Gamma (D \rightarrow V_1 P_2) &= {1 \over 8 \pi}\ {k^3 \over m^2_1}\ |B|^2
&(11)\cr
\noalign{\vskip 4pt}%
\Gamma (D \rightarrow V_1 V_2) &= {1 \over 8 \pi}\ {k \over M^2}\
\bigg[ |C|^2 \bigg( 3 + {M^2 k^2 \over m^2_1 m^2_2} \bigg)\cr
 &- (CE^* + C^* E) \ {M^2 (M^2 - m^2_1 - m^2_2) \over 2 m^2_1 m^2_2} \ k^2\cr
\noalign{\vskip 4pt}%
& +
|D|^2\  2\ M^2 k^2 + |E|^2 \
{M^4 k^4 \over m^2_1 m^2_2} \bigg] &(12)\cr}$$
where
$$k = {1 \over 2M}\ \left[ \left( M^2 - m^2_1 - m^2_2 \right)^2 - 4\
m^2_1 m^2_2 \right]^{1/2} \quad , \eqno(13)$$
$M$ is the mass of the decaying particle and $m_1$, $m_2$ are the mases of
the particles in the decay product.  The amplitudes $A,\  B \dots E$ can be
written down from the Feynman diagrams in Figs. 2-4.  Using the numerical
values of the couplings in Table 1, these can be expressed in terms of the
four parameters: $a_1,$ $a_2$, $g_s$ and $g_d$.  The results for the
Cabibbo allowed and once-suppressed decays are listed$^{17}$ in Tables 2-4.

\medskip

\noindent {\bf 4. \underbar{Results}}

\smallskip

The decay amplitudes for $D \rightarrow PP$ listed in Table 2, do not
involve the parameters $g_s$ and $g_d$.  Thus it is best to obtain a fit
for $a_1$ and $a_2$ from these decays.  For this purpose we shall use the
data on $D \rightarrow K \pi$ decays.

So far we have ignored the final state interactions.  We shall take these
into account by considering only elastic scattering in the final state.  In
$D \rightarrow K \pi$ decays, we have in terms of isospin amplitudes
$$\eqalign{A (D^0 \rightarrow K^- \pi^+) &= {1 \over \sqrt{3}} \
A_{3/2} + \sqrt{{2 \over 3}} \ A_{1/2}\cr
A (D^0 \rightarrow \overline K^0 \pi^0) &=
  \sqrt{{2 \over 3}} \ A_{3/2} - {1 \over \sqrt{3}} \
A_{1/2}\cr
A (D^+ \rightarrow \overline K^0 \pi^+) &=
\sqrt{3}\ A_{3/2}\cr}\eqno(14)$$
where
$$A_I = | A_I | e^{i \delta_I} $$
is the amplitude in the isospin state $I$ and $\delta_I$ is the phase shift
in that channel.  Using the data from the particle properties data
booklet,$^{18}$
it is easy to obtain
$$\eqalign{&| A_{1/2}| = 2.94 \times 10^{-6} \ {\rm GeV}\cr
&|A_{3/2}| =  7.37 \times 10^{-7} \ {\rm GeV}\cr
&\delta_{1/2} - \delta_{3/2} = 93.4^\circ\cr}\eqno(15)$$
Now, in our model, the isospin amplitudes can be constructed if we invert
the relations in Eq. (14) and use our results for the $D \rightarrow K \pi$
amplitudes from Table 2.  This gives
$$\eqalign{&|A_{1/2}| = 1.18 \times 10^{-6} (a_1 - 1.04 a_2)\
{\rm GeV}\cr
&|A_{3/2}| = 8.37 \times 10^{-7} (a_1 + 1.35 a_2)\ {\rm GeV}\cr}\eqno(16)$$
Using the values for these amplitudes given by (15), we obtain the
solution$^{19}$
$$a_1 = 1.79 \quad , \quad a_2 = -0.67 \eqno(17)$$
There is another solution where $|a_2 / a_1 | > 1$, which we discard, as
discussed in ref. (3). We also note that
the result (17) is not very far from the values of
$a_1$ and $a_2$ obtained by Bauer et al.$^4$

With $a_1$ and $a_2$ determined, we fit $g_s$ and $g_d$ from the data on
the decay modes $D \rightarrow PV$ and $D \rightarrow VV$.  We find that
the best fit corresponds to the values
$$g_s = - 10.14 \quad , \quad g_d = -9.85 \eqno(18)$$
The branching ratios for various decay modes determined by our values of the
parameters in (17) and (18) are exhibited in Tables 5-7, together with the
experimental data.  In preparing these tables, we have used, wherever
possible, phase shifts for elastic scattering in the final state as
determined from the data.  Considering the uncertainties in our choice of
the coupling constants,
 it is remarkable that most of
 our results are in reasonable agreement with the
data.

It should be remarked that our parameter fit in (17) and (18) leads to a
destructive interference between the pseudoscalar and axial-vector meson poles
in the annihilation diagrams in $D \rightarrow PV$.  Thus decays like
$D^0 \rightarrow \phi \overline K^0$ and
$D^+_S \rightarrow \rho^+ \pi^0$ which proceed only through the
annihilation diagrams lead to tiny branching ratios due to large
cancellations in the contributions of figures (3c) and (3e).
Unfortunately, because of this, small changes in our choice of the
couplings can get magnified in the prediction of these branching
ratios.$^{20}$
We would also like to point out that in $D \rightarrow VV$ decays, we did
not consider the diagram analogous to (4c) where the pseudoscalar meson
pole is replaced by an axial-vector meson.
There is no straightforward way of estimating the $AVV$ strong coupling.
The general agreement of our results
 with experiments
 in fact provides a justification for  the
neglect of the axial-vector pole diagram in $D \rightarrow VV$.
\vfill\eject
\vskip 1cm
\centerline{\bf \underbar{Table 1}}
\midinsert
\thicksize=.5pt
\thinsize=.5pt
\tablewidth=4in
\begintable
Quantity | Value \crnorule
| \cr
$g$ | 4.28 \cr
$g_V/m_V$ | 152 MeV \cr
$f_\pi$ | 92.6 MeV \cr
$f_K$ | 1.2 $f_\pi$ \cr
$f_\eta$ | $f_\pi$ \cr
$f_{\eta^\prime}$ | $f_\pi$ \cr
$f_D$ | 1.46 $f_\pi$ \cr
$f_{D_S}$ | $f_D$ \cr
$g_{A_1} / m_{A_1}$ | 121 MeV \cr
$g_{K_1} /m_{K_1}$ | 104 MeV \cr
$g_{f_1} / m_{f_1}$ | 121 MeV \cr
$g_{D_1} / m_{D_1}$ | 69.5 MeV \cr
$g_{D_{S1}} / m_{D_{S1}}$ | 69.5 MeV \endtable
\vskip .1in
\endinsert\noindent
\vfill\eject

\vskip 1cm
\centerline{\bf \underbar{Table 2}}
\midinsert
\thicksize=.5pt
\thinsize=.5pt
\tablewidth=4in
\begintable
Decay | Amplitude $A(\times 10^6$ Gev$^{-1})$ \crnorule
| \cr
$D^0 \rightarrow K^- \pi^+$ | 1.45 ($a_1 - 0.24\  a_2$) \cr
$D^0 \rightarrow \overline K^0 \pi^0$ | 1.63 $a_2$ \cr
$D^0 \rightarrow \overline K^0 \eta$ | 0.50 $a_2$\cr
$D^+ \rightarrow \overline K^0 \pi^+$ | 1.45 ($a_1 + 1.35\  a_2$)\cr
$D^+_S \rightarrow \overline K^0 K^+$ | 2.05 $a_2$ \cr
$D^+_S \rightarrow \eta \pi^+$ | $-0.99 \ a_1$\cr
$D^+_S \rightarrow \eta^\prime \pi^+$ | 1.75 $a_1$ \cr
| \cr
$D^0 \rightarrow K^+ K^-$ | $- 0.39\  a_1$\cr
$D^0 \rightarrow K^0 \overline K^0$ | 0\cr
$D^0 \rightarrow \pi^+ \pi^-$ | $-$0.37 $a_1$\cr
$D^0 \rightarrow \pi^0 \pi^0$ | $-$0.37 $a_2$\cr
$D^+ \rightarrow K^+ \overline K^0$ | $-$0.39 $a_1$\cr
$D^+ \rightarrow \pi^+ \pi^0$ | $-$ 0.26 $(a_1 + a_2)$ \cr
$D^+ \rightarrow \eta \pi^+$ | 0.17 $a_2 + 0.21\ a_1$\cr
$D^+ \rightarrow \eta^\prime \pi^+$ | 0.08 $a_2 +$ 0.36 $a_1$\cr
$D^+_S \rightarrow K^0 \pi^+$ | $0.47\ a_1$\endtable
\vskip .1in
\endinsert\noindent
\vfill\eject
\vsize= 9.5 truein
\centerline{\bf \underbar{Table 3}}
\midinsert
\thicksize=.5pt
\thinsize=.5pt
\tablewidth=6in
\begintable
Decay | Amplitude $B\ (\times 10^6)$\cr
$D^0 \rightarrow K^{* -} \pi^+$ | $2.16\ a_2 + 0.008\ a_1 +
0.323\ a_2 (g_s - 0.611 g_d) - 0.074\ a_1 (g_s +
0.382 g_d)$ \cr
$D^0 \rightarrow \overline K^{* 0} \pi^0$ | $- 0.320\ a_2 - 0.230\ a_2
(g_s - 0.611 g_d)$\cr
$D^0 \rightarrow \rho^+ K^-$ | $-2.16\ a_2 + 1.30\ a_1 -
0.323\ a_2 (g_s - 0.548 g_d)$\cr
$D^0 \rightarrow \rho^0 \overline K^0$ | $1.64\ a_2 + 0.230\ a_2
(g_s - 0.548 g_d) - 0.066\ a_2 (g_s +
0.379 g_d)$ \cr
$D^0 \rightarrow \phi \overline K^0$ | $2.16\ a_2 + 0.323\ a_2 (g_s -
 0.614 g_d)$\cr
$D^0 \rightarrow \omega \overline K^0$ | $- 1.43\ a_2 - 0.230\
a_2 (g_s - 0.554 g_d) - 0.066\ a_2 (g_s + 0.376 g_d)$\cr
$D^0 \rightarrow \overline K^{* 0}\eta$ | $3.67\ a_2 + 0.388\ a_2
(g_s - 0.571 g_d)$\cr
$D^+ \rightarrow \overline K^{* 0} \pi^+$ | $1.72\ a_2 + 0.008\ a_1 -
0.074\ a_1 (g_s + 0.382 g_d)$\cr
$D^+ \rightarrow \rho^+ \overline K^0$ | $0.153\ a_2 + 1.30\ a_1 -
0.093\ a_2 (g_s + 0.379 g_d)$\cr
$D^+_S \rightarrow \rho^+ \pi^0$ | $-2.38\ a_1 - 0.565\ a_1
(g_s - 0.573 g_d)$\cr
$D^+_S \rightarrow K^{* +} \overline K^0$
 | $1.69\ a_1 + 0.153\ a_2 +
0.400\ a_1 (g_s - 0.570 g_d) - 0.098\ a_2 (g_s +
0.366 g_d)$ \cr
$D^+_S \rightarrow \overline K^{* 0} K^+$ | $-1.56\ a_1 + 1.53\ a_2 -
0.400\ a_1 (g_s - 0.570 g_d)$\cr
$D^+_S \rightarrow \phi \pi^+$  | $0.008 \ a_1 - 0.078\ a_1
(g_s + 0.364 g_d)$\cr
$D^+_S \rightarrow \omega \pi^+$ | 0\cr
$D^+_S \rightarrow \rho^+ \eta$ | $-0.756 \ a_1$\cr
$D^+_S \rightarrow \rho^+ \eta^\prime$ | $1.68\ a_1$ \cr
| \cr
$D^0 \rightarrow K^{* -}K^+$  | $0.355\ a_2 + 0.031\ a_1 +
0.085\ a_2 (g_s - 0.580 g_d) - 0.020\ a_1 (g_s +
0.350 g_d)$ \cr
$D^0 \rightarrow K^{* +}K^-$ | $-0.355\ a_2 + 0.361\ a_1 -
0.085\ a_2 (g_s - 0.580 g_d)$\cr
$D^0 \rightarrow \overline K^{* 0}K^0$ | $-0.029\ a_2 - 0.001\ a_2
(g_s - 0.580 g_d)$\cr
$D^0 \rightarrow K^{* 0} \overline K^0$ | $0.029\ a_2 + 0.001\ a_2
(g_s - 0.580 g_d)$\cr
$D^+ \rightarrow \overline K^{* 0}K^+$ | $-0.353\ a_1 - 0.086\ a_1
(g_s - 0.580 g_d) - 0.020\ a_1 (g_s +
0.350 g_d)$ \cr
$D^+ \rightarrow \phi \pi^+$ | $0.479\ a_2$\cr
$D^+ \rightarrow \omega \pi^+$ | $0.001\ a_1 + 0.228\ a_2 -
0.012\ a_1 (g_s + 0.409 g_d)$\cr
$D^+ \rightarrow \rho^0 \pi^+$ | $0.541\ a_1 - 0.223\ a_2 +
0.122\ a_1 (g_s - 0.582 g_d) + 0.012\ a_1 (g_s +
0.413 g_d)$ \cr
$D^+ \rightarrow \rho^+ \eta$  | $-0.010\ a_2 + 0.196\ a_1 +
0.005\ a_2 (g_s + 0.373 g_d)$\cr
$D^+ \rightarrow \rho^+ \eta^\prime$ | $0.087\ a_2 + 0.334\ a_1 -
0.011\ a_2 (g_s + 0.284 g_d)$\endtable
\vskip .1in
\endinsert\noindent
\vfill\eject

\centerline{\bf \underbar{Table 4}}
\midinsert
\thicksize=.5pt
\thinsize=.5pt
\tablewidth=6in
\begintable
Decay | &Amplitudes&\cr
  | $C\ (\times 10^6 \ {\rm GeV}^{-1})$ | $D\ (\times 10^6 \ {\rm GeV})$
| $E\ (\times 10^6 \ {\rm GeV})$\cr
$D^0 \rightarrow K^{* -} \rho^+$ | $-0.215\ a_2 - 0.103 \ a_1 g_s$ |
$-1.16\ a_2 - 0.696\ a_1$ | 0.030 $a_1 g_d$\cr
$D^0 \rightarrow \overline K^{* 0} \rho^0$ | $0.152\ a_2 - 0.093 \ a_2 g_s$ |
0.171 $a_2$ | 0.027 $a_2 g_d$\cr
$D^0 \rightarrow \overline K^{* 0} \omega$ | $-0.136\ a_2 - 0.093 \ a_2 g_s$ |
$-1.47\ a_2$ | 0.027 $a_2 g_d$\cr
$D^+ \rightarrow \overline K^{* 0} \rho^+$ | $-0.103\ a_1
 g_s  - 0.131 \ a_2 g_s$ |
$-0.924\ a_2 - 0.696\ a_1$ | 0.030 $a_1 g_d + 0.038\ a_2 g_d$\cr
$D^+_S \rightarrow \phi \rho^+$ | $-0.108\ a_1 g_s$  |
$-0.660\ a_1$ | 0.028 $a_1 g_d$\cr
$D^+_S \rightarrow K^{* +} \overline K^{*0}$ | $-0.139\ a_2 g_s$|
$-0.873\ a_2 - 0.853\ a_1$ | 0.036 $a_2 g_d$\cr
| | | \cr
$D^0 \rightarrow K^{* 0} \overline K^{*0}$ | 0 |
 $- 0.241 \ a_2$ | 0 \cr
$D^0 \rightarrow \phi \rho^0 $ | $-0.025\ a_2 g_s$|
$-0.181\ a_2$ | 0.007 $a_2 g_d$\cr
$D^+ \rightarrow K^{* +} \overline K^{*0}$ | $-0.028\ a_1 g_s$|
$-0.400\ a_1$ | 0.008 $a_1 g_d$\cr
$D^+ \rightarrow \phi \rho^+$ | $-0.036\ a_2 g_s$|
$-0.256\ a_2$ | 0.010 $a_2 g_d$\endtable
\vskip .1in
\endinsert\noindent
\vfill\eject

$$
\table
\tablespec{\c\c\c}
\spacing{18pt}
\body{
\header{\underbar{Table 5}}
\skip{5pt}
\hline
| \extend1{Decay} | \extend2{Branching Ratio}|\end
\hdash{0--}
|    |Theory & Experiment |\end
\hline
| | | |\end
|$D^0 \rightarrow K^- \pi^+$ | $3.6 \times 10^{-2}$ |
$(3.65 \pm 0.21) \times 10^{-2}$|\end
\hline
|$D^0 \rightarrow \overline K^0 \pi^0$ | $2.1 \times 10^{-2}$ |
$(2.1 \pm 0.5) \times 10^{-2}$|\end
\hline
|$D^0 \rightarrow \overline K^0 \eta$ | $6.4 \times 10^{-4}$ |
$< 2.3 \times 10^{-2}$|\end
\hline
|$D^+ \rightarrow \overline K^0 \pi^+$ | $2.6 \times 10^{-2}$ |
$(2.6 \pm 0.4) \times 10^{-2}$|\end
\hline
|$D^+_S \rightarrow \overline K^0 K^+$ | $1.1 \times 10^{-2}$ |
$(2.8 \pm 0.7) \times 10^{-2}$|\end
\hline
|$D^+_S \rightarrow \eta \pi^+$ | $2.0 \times 10^{-2}$ |
$(1.5 \pm 0.4) \times 10^{-2}$|\end
\hline
|$D^+_S \rightarrow \eta^\prime \pi^+$ | $5.1 \times 10^{-2}$ |
$(3.7 \pm 1.2) \times 10^{-2}$|\end
\hline
| | | | \end
\hline
|$D^0 \rightarrow K^+ K^-$ | $2.4 \times 10^{-3}$ |
$(4.1 \pm 0.4) \times 10^{-3}$|\end
\hline
|$D^0 \rightarrow K^0 \overline K^0$ | $5.0 \times 10^{-4}$ |
$(1.1 \pm 0.4) \times 10^{-3}$|\end
\hline
|$D^0 \rightarrow \pi^+ \pi^-$ | $1.6 \times 10^{-3}$ |
$(1.63 \pm 0.19) \times 10^{-3}$|\end
\hline
|$D^0 \rightarrow \pi^0 \pi^0$ | $3.9 \times 10^{-5}$ |
$< 4.6  \times 10^{-3}$|\end
\hline
|$D^+ \rightarrow K^+ \overline K^0$ | $7.3 \times 10^{-3}$ |
$(7.3 \pm 1.8) \times 10^{-3}$|\end
\hline
|$D^+ \rightarrow \pi^+ \pi^0$ | $1.5 \times 10^{-3}$ |
$< 5.3  \times 10^{-3}$|\end
\hline
|$D^+ \rightarrow \eta \pi^+$ | $1.1 \times 10^{-3}$ |
$(6.6 \pm 2.2) \times 10^{-3}$|\end
\hline
|$D^+ \rightarrow \eta^\prime \pi^+$ | $4.4 \times 10^{-3}$ |
$< 8  \times 10^{-3}$|\end
\hline
|$D^+_S \rightarrow K^0 \pi^+$ | $4.5 \times 10^{-3}$ |
$< 6  \times 10^{-3}$|\end
\hline
}
\endtable
$$
\vfill\eject

$$
\table
\tablespec{\c\c\c}
\spacing{18pt}
\body{
\header{\underbar{Table 6}}
\skip{5pt}
\hline
| \extend1{Decay} | \extend2{Branching Ratio}|\end
\hdash{0--}
|    |Theory | Experiment |\end
\hline
%| | | |\end
%\hline
|$D^0 \rightarrow K^{*-} \pi^+$ | $1.4 \times 10^{-2}$ |
$(4.5 \pm 0.6) \times 10^{-2}$|\end
\hline
|$D^0 \rightarrow \overline K^{*0} \pi^0$ | $6.8 \times 10^{-3}$ |
$(2.1 \pm 1.0) \times 10^{-2}$|\end
\hline
|$D^0 \rightarrow \rho^+ K^-$ | $10.2 \times 10^{-2}$ |
$(7.3 \pm 1.1) \times 10^{-2}$|\end
\hline
|$D^0 \rightarrow \rho^0 \overline K^0$ | $1.3 \times 10^{-2}$ |
$(6.1 \pm 3.0) \times 10^{-3}$|\end
\hline
|$D^0 \rightarrow \phi \overline K^0$ | $1.1 \times 10^{-3}$ |
$(8.8 \pm 1.2) \times 10^{-3}$|\end
\hline
|$D^0 \rightarrow \omega \overline K^0$ | $1.8 \times 10^{-3}$ |
$(2.5 \pm 0.5) \times 10^{-2}$|\end
\hline
|$D^0 \rightarrow \overline K^{*0} \eta$ | $1.0 \times 10^{-2}$ |
$(2.1 \pm 1.2) \times 10^{-2}$|\end
\hline
|$D^+ \rightarrow \overline K^{*0} \pi^+$ | $1.4 \times 10^{-2}$ |
$(1.9 \pm 0.7) \times 10^{-2}$|\end
\hline
|$D^+ \rightarrow \rho^+ \overline K^0$ | $6.4 \times 10^{-2}$ |
$(6.6 \pm 1.7) \times 10^{-2}$|\end
\hline
|$D^+_S \rightarrow \rho^+ \pi^0$ | $2.2 \times 10^{-3}$ |
$< 2.2  \times 10^{-3}$|\end
\hline
|$D^+_S \rightarrow K^{*+} \overline K^0$ | $1.6 \times 10^{-2}$ |
$(3.3 \pm 0.9)  \times 10^{-2}$|\end
\hline
|$D^+_S \rightarrow \overline K^{*0} K^+$
 | $3.6 \times 10^{-3}$ |
$( 2.6  \pm 0.5) \times 10^{-2}$|\end
\hline
|$D^+_S \rightarrow \phi \pi^+ $
 | $3.5 \times 10^{-2}$ |
$( 2.8  \pm 0.5) \times 10^{-2}$|\end
\hline
|$D^+_S \rightarrow \omega \pi^+$
 | 0 |
$< 1.4   \times 10^{-2}$|\end
\hline
|$D^+_S \rightarrow \rho^+ \eta$
 | $3.3 \times 10^{-2}$ |
$(7.9  \pm 2.1) \times 10^{-2}$|\end
\hline
|$D^+_S \rightarrow \rho^+ \eta^\prime$
 | $4.4 \times 10^{-2}$ |
$(9.5  \pm 2.7) \times 10^{-2}$|\end
\hline
| | | |\end
\hline
|$D^0 \rightarrow  K^{*-} K^+$ | $2.2 \times 10^{-3}$ |
$(2.0 \pm 1.0) \times 10^{-3}$|\end
\hline
|$D^0 \rightarrow  K^{*+} K^-$ | $2.9 \times 10^{-3}$ |
$(3.5 \pm 0.8) \times 10^{-3}$|\end
\hline
|$D^0 \rightarrow  \overline K^{*0} K^0$ | $2.0 \times 10^{-6}$ |
$< 1.6  \times 10^{-3}$|\end
\hline
|$D^0 \rightarrow  K^{*0} \overline K^0$ | $2.0 \times 10^{-6}$ |
$< 8 \times 10^{-4}$|\end
\hline
|$D^+ \rightarrow  \overline K^{*0} K^+$ | $5.3 \times 10^{-3}$ |
$(4.7 \pm 0.9) \times 10^{-3}$|\end
\hline
|$D^+ \rightarrow  \phi \pi^+$ | $1.7 \times 10^{-3}$ |
$(6.0 \pm 0.8) \times 10^{-3}$|\end
\hline
|$D^+ \rightarrow  \omega \pi^+$ | $1.1 \times 10^{-3}$ |
$< 6  \times 10^{-3}$|\end
\hline
|$D^+ \rightarrow  \rho^0 \pi^+$ | $1.1 \times 10^{-3}$ |
$< 1.2  \times 10^{-3}$|\end
\hline
|$D^+ \rightarrow  \rho^+ \eta$ | $5.1 \times 10^{-3}$ |
$< 1.0  \times 10^{-2}$|\end
\hline
|$D^+ \rightarrow  \rho^+ \eta^\prime$ | $9.6 \times 10^{-4}$ |
$< 1.4  \times 10^{-2}$|\end
\hline
}
\endtable
$$
\vfill\eject

$$
\table
\tablespec{\c\c\c}
\spacing{18pt}
\body{
\header{\underbar{Table 7}}
\skip{5pt}
\hline
| \extend1{Decay} | \extend2{Branching Ratio}|\end
\hdash{0--}
|    |Theory & Experiment |\end
\hline
| | | |\end
|$D^0 \rightarrow K^{*-} \rho^+$ | $6.5 \times 10^{-2}$ |
$(6.2 \pm 2.5) \times 10^{-2}$|\end
\hline
|$D^0 \rightarrow \overline K^{*0} \rho^0$ | $8.5 \times 10^{-3}$ |
$(1.5 \pm 0.6) \times 10^{-2}$|\end
\hline
|$D^0 \rightarrow \overline K^{*0} \omega$ | $7.9 \times 10^{-3}$ |
$< 1.5  \times 10^{-2}$|\end
\hline
|$D^+ \rightarrow \overline K^{*0} \rho^+$ | $4.3 \times 10^{-2}$ |
$\big( 4.1  { + 1.5 \atop -1.2}\big) \times 10^{-2}$|\end
\hline
|$D^+_S \rightarrow \phi \rho^+$ | $5.7 \times 10^{-2}$ |
$\big( 5.2 {+ 1.4 \atop -1.6} \big) \times 10^{-2}$|\end
\hline
|$D^+_S \rightarrow K^{*+} \overline K^{*0}$ | $1.5 \times 10^{-2}$ |
$(5.0 \pm 1.7) \times   10^{-2}$|\end
\hline
| | | | \end
\hline
|$D^0 \rightarrow K^{*0} \overline K^{*0}$ | $2.6 \times 10^{-5}$ |
$\big( 2.7 {+1.5 \atop  - 1.2} \big) \times 10^{-3}$|\end
\hline
|$D^0 \rightarrow \phi  \rho^0$ | $2.2 \times 10^{-4}$ |
$(1.8 \pm 0.5) \times 10^{-3}$|\end
\hline
|$D^+ \rightarrow K^{*+} \overline K^{*0}$ | $5.9 \times 10^{-3}$ |
$(2.6 \pm 1.1) \times 10^{-2}$|\end
\hline
|$D^+ \rightarrow \phi  \rho^+$ | $1.2 \times 10^{-3}$ |
$< 1.3  \times 10^{-2}$|\end
\hline
}
\endtable
$$

\vfill\eject
\noindent {\bf \underbar{References}}
\smallskip

\item{1.} V. S. Mathur, Proceedings of a Conference on Quark Confinement
and Field Theory, Rochester, NY (1976), edited by D. R. Stump and D. H.
Weingarten (published by John Wiley \& Sons).

\item{2.} J. J. Sakurai, Phys. Rev. {\bf 156}, 1508 (1967).

\item{3.} Ashok Das and Vishnu S. Mathur, Mod. Phys. Lett A (to be
published).

\item{4.} M. Bauer, B. Stech and M. Wirbel, Z. Phys. C -- Particles and
Fields {\bf 34}, 103 (1987).

\item{5.} For a  description
of other methods used in the study of charm decay, see the review papers:
M. Wirbel, Prog. in Particle and Nuclear Phys. {\bf 21}, 33 (1988); R. J.
Morrison and M. S. Witherell, Annual Rev. of Nuclear and Particle Science
{\bf 39}, 183 (1989).

\item{6.} K. Jagannathan and V. S. Mathur, Nucl. Phys. {\bf B171}, 78
 (1980) and ref. 4.

\item{7.} B. Sakita and K. C. Wali, Phys. Rev. Lett. {\bf 14}, 404
(1965) and Phys. Rev. {\bf 139}, B1355 (1965); A. Salaam, R. Delbourgo and
J. Strathdee, Proc. Roy. Soc. (London) {\bf 284}, 146 (1965).

\item{8.} T. Das, V. S. Mathur and S. Okubo, Phys. Rev. Lett. {\bf 19},
1067
 (1967);  H. J. Schnitzer and S. Weinberg, Phys. Rev. {\bf 164}, 1828
(1967).

\item{9.} V. S. Mathur, Phys. Rev. {\bf 174}, 1743 (1968).

\item{10.} S. Weinberg, Phys. Rev. Lett. {\bf 18}, 507 (1967).

\item{11.}  T. Das, V. S. Mathur and S. Okubo, Phys. Rev. Lett.
{\bf 18}, 761 (1967).  In the present paper, we use only the
pole-dominated form of the first spectral-function sum-rule.

\item{12.} T. Das, V. S. Mathur and S. Okubo, Phys. Rev. Lett. {\bf 19},
470 (1967).

\item{13.} This is a generalization to chiral SU(4) of the original
(pole-dominated) first spectral-function sum-rule of Weinberg (ref. 10 and
11).  Note, in passing, that (6) leads to the upper bound
$f_P < g_V / m_V = 152$ MeV.

\item{14.} We use the value determined in V. S. Mathur and M. T. Yamawaki,
Phys. Rev. {\bf D29}, 2057 (1984).  See also S. Narison, Phys. Lett.
{\bf 198B}, 104 (1987) and C. A. Dominguez and N. Paver, Phys. Lett.
{\bf 197B}, 423 (1987).

\item{15.} M. Shifman, A. Vainshtein and V. Zakharov, Nucl. Phys.
{\bf B147}, 385 (1979); {\bf B147}, 448 (1979).

\item{16.} D. W. G. S. Leith in Proceedings of the 6th NATO Advanced Study
Institute on Techniques and Concepts of High-Energy Physics VI edited by
Thomas Ferbel and published by Pelnum Press, New York (1991).

\item{17.} Only those decay modes have been listed for which there is some
experimental data.

\item{18.} Review of Particle Properties, Phys. Rev. {\bf D45}, Part 2
(June 1992).  We have ignored the errors quoted in the data for our
analysis here.

\item{19.} These values of $a_1$ and $a_2$ are slightly different from
those in ref. 3 since we have used a slightly different value of the strong
coupling constant $g$.

\item{20.} In fact  the fit (18) for $g_s$ and $g_d$ was fine-tuned to
reproduce the experimental bound for the decay $D^+_S \rightarrow
 \rho^+ \pi^0$.

\vfill\eject

\noindent {\bf \underbar{Table Captions}}

\medskip

\item{Table 1.} Numerical values of the coupling constants and other
parameters used in the paper.

\item{Table 2.} Amplitude $A$ defined in Eq. (7) for the decay
 $D \rightarrow P_1 P_2$.

\item{Table 3.} Amplitude $B$ defined in Eq. (8) for the decay
$D \rightarrow V_1 P_2$.

\item{Table 4.} Amplitudes $C$, $D$ and $E$ defined in Eq. (9)
 for the decay $D \rightarrow V_1 V_2$.

\item{Table 5.} Branching ratio for the decay $D \rightarrow P_1 P_2$.

\item{Table 6.} Branching ratio for the decay $D \rightarrow V_1 P_2$.

\item{Table 7.} Branching ratio for the decay $D \rightarrow V_1 V_2$.

\bigskip
\bigskip

\noindent {\bf \underbar{Figure Captions}}

\medskip

\item{Fig. 1.} Quark-model diagrams for the nonleptonic decay of $D$ into
two mesons.  1(a) describe the spectator diagrams and 1(b) the annihilation
diagrams.

\item{Fig. 2.} Feynman diagrams for the decay $D \rightarrow P_1 P_2$.

\item{Fig. 3.} Feynman diagrams for the decay $D \rightarrow V_1 P_2$.

\item{Fig. 4.} Feynman diagrams for the decay $D \rightarrow V_1 V_2$.

\end

\end